\documentclass[%
 reprint,
superscriptaddress,
 amsmath,amssymb,
 aps,
prb,
floatfix,
]{revtex4-2}
\usepackage[ngerman,english]{babel}
\usepackage{graphicx}
\usepackage{braket}
\usepackage[percent]{overpic}
\usepackage{graphicx}
\usepackage{dcolumn}
\usepackage{bm}
\usepackage{comment}
\usepackage{xcolor}
\usepackage[separate-uncertainty=true]{siunitx}
\usepackage[normalem]{ulem}
\usepackage{subfigure}
\usepackage{printlen}
\uselengthunit{pt}
\usepackage[colorlinks=true,linkcolor=blue,citecolor=blue,urlcolor=blue]{hyperref}

\begin{document}
\author{Lefan Dolg}
\email{lefan.zhang@rwth-aachen.de}
\affiliation{Institute for Theoretical Solid State Physics, RWTH Aachen, Germany}
\affiliation{Institute for Theory of Statistical Physics, RWTH Aachen, Germany}
\author{Moritz Scharfstädt}
\affiliation{Physikalisches Institut, University of Bonn, Germany}
\author{Andrea Bergschneider}
\affiliation{Physikalisches Institut, University of Bonn, Germany}
\author{Dante M. Kennes}
\affiliation{Institute for Theory of Statistical Physics, RWTH Aachen, Germany}
\affiliation{Max Planck Institute for the Structure and Dynamics of Matter, Center for Free Electron Laser Science Hamburg, Germany}
\author{Silvia {Viola Kusminskiy}}
\affiliation{Institute for Theoretical Solid State Physics, RWTH Aachen, Germany}
\affiliation{Max Planck Institute for the Science of Light, Erlangen, Germany}

\date{\today}
\title{Numerical investigation of electrostatically confined excitons in monolayer $\text{MoSe}_2$}

\begin{abstract}
    We investigate exciton confinement to a quantum wire in monolayer $\text{MoSe}_2$ where the confinement is achieved by a p-i-n junction. We employ an effective-mass exciton model and solve the problem numerically, reflecting device geometries found in experimental state-of-the-art set ups. Our method allows us to investigate the entire spectrum of confined states. 
    We show the emergence of quantum confinement and study the dependence of the confined states as a function of electrical gate voltages, which are experimentally tunable parameters. We find that the confined states can be divided into bright and dark states with the dark states having small but finite oscillator strengths. Their oscillator strengths are low enough that they have not yet been detected in experiments, whereas the spectrum of the bright exciton states reproduces recent experimental measurements.
    Our results provide insight into the theoretical background of confined exciton states beyond the ground state and pave the way for the development of new confinement schemes as well as avenues to access the previously not detected dark states.
\end{abstract}

\maketitle

\section{\label{sec:label} Introduction}

Transition-metal dichalcogenides (TMD) are a class of materials that harbor a wide range of interesting physical phenomena, including tunable bandgaps and significant spin-orbit coupling ~\cite{Quintela2022, Manzeli2017, Cheiwchanchamnangij2012}.
The crystal structure of bulk TMDs consists of atomically thin layers held together by van der Waals interactions~\cite{KOLOBOV2018}. Therefore TMD monolayers can be obtained by the same exfoliation methods as graphene \,\cite{KOLOBOV2018, CastellanosGomez2014, Backes2016} making them a prime candidate for studying 2D physics. 
The transition from bulk to monolayers in TMDs changes the indirect bandgap at the $\mathbf{K^+}/\mathbf{K^-}$ points of the Brillouin zone (BZ) into direct bandgaps\,\cite{Quintela2022,Arora2015}, which leads to strong light-matter coupling~\cite{Arora2015,Wang2018,LopezSanchez2013,zhang2014,pollmann2015, Jones2013}. 
The optical response of TMDs is dominated by exciton peaks \cite{Raja2017, Cadiz2017, Qiu2015}. Excitons are quasi-particle bound states created by excited electrons in the conduction band and electron-holes in the valence band.
In monolayers, exciton binding energies are increased due to the reduced screening of the Coulomb interaction \,\cite{Wang2018}. 
The strong light-matter interaction combined with the large binding energies makes excitons in monolayers an excellent target for the development of opto-electronic interfaces \,\cite{LopezSanchez2013}. 
The electronic properties of these excitons have therefore been extensively studied during the past decade~\cite{Glazov2018,Wegerhoff2025,Dark_excitons,Li2020,Bieniek2022,Exciton_band, Glazov2014,dery2025,Efekim2021,Li2014,Ruiz2020,Yu2015}.

\begin{figure}[htpb]
    \centering
    \includegraphics[width = \linewidth]{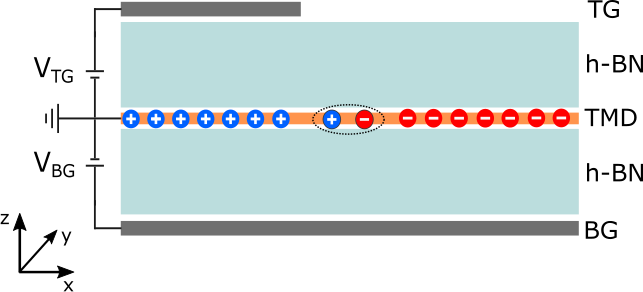}
    \caption{ Schematic representation of the simulated experimental set-up. The TMD monolayer is sandwiched between two layers of h-BN and electrical gates on top (TG) and bottom (BG). Different bias voltages $V_{\text{TG}},V_{\text{BG}}$ applied to the gates combined with the asymmetric spatial extent of the electric gates creates an electrical potential difference in x-direction.}
    \label{fig:set_up}
\end{figure}
Confining these excitons below their De-Broglie wavelength results in discrete motional states and energy levels. Being able to manipulate the exciton confinement length leads to control over the exciton spectrum and therefore also grants control over the light-matter interaction. 
However the charge neutrality of excitons makes it intrinsically hard to confine them. Possible confinement approaches include the embedding of defects in TMD monolayers~\cite{Kim2024,Qu2024} and by an in-plane electric field through the DC-Stark effect. The latter is easily tunable through the field strength. First computational approaches~\cite{BOA,excitonlocal} have shown that the ground-state properties of the confined exciton, such as its radius and position, can be tuned by the strength of the electric field. 
The first experimental realization of one-dimensional (1D) confinement was reported by Thureja et al.~\cite{Thureja2022}. More recently, zero-dimensional (0D) confinement in a quantum dot was demonstrated independently by two groups~\cite{thureja2024,Hu2024}. In this work we will focus on the 1D confinement case. Experimentally, the confinement is achieved by embedding the sample in electrical gates with different spatial extent, see Fig.~\ref{fig:set_up}. An in-plane potential difference is then induced by applying different voltage biases to the gates. Additionally, the voltage in the gates induce charge doping, which leads to positively and negatively doped areas in the sample. This kind of gate setup is therefore also called a p-i-n junction\,\cite{excitonlocal}. 


While the ground state of the confined exciton has been studied~\cite{BOA,excitonlocal}, 
measurements in experimental realizations~\cite{Thureja2022,Heithoff} also show the emergence of excited states. 
For applications in opto-electronic devices, investigating the entire spectrum of confined states is of interest. Understanding the intricacies of the excited states and their entire resonance spectrum helps to unlock the full potential of an exciton based device. Furthermore, the potentials used in the previous computational studies~\cite{BOA,excitonlocal} utilize symmetric analytic functions to approximate the confining potential. However, the confining potentials in experiments deviate from this assumption and are heavily influenced by screening effects resulting from the buildup of doping charge. Finding simple analytic functions that approximate the potential in an experimental sample over the entire parameter range is difficult. A theoretical study with realistic potentials obtained from electrostatic simulations can provide direct comparison to experimental results and serve to accurately guide the confinement design.

In this work we perform a full computational study of exciton confinement including the excited states,  using confining potentials closely emulating the experimental setup. 
We find that the oscillator strengths of the confined states are mainly dependent on the parity of their wavefunction in the center of mass coordinate. The parity of the wavefunction divides the states into dark and bright states with the dark states carrying oscillator strengths that are one order of magnitude lower than their bright counterpart. We find that the small but finite oscillator strength in the dark states is caused by the asymmetrical confining potential. Additionally we observe that the oscillator strength is the strongest for the confined states right after they emerge out of the 2D limit.  
Our results show that at large voltage biases at the electric gates, the build up of doping charge screens the confining potential causing a convergence behavior in the potential. Hence, going beyond the experimentally available parameter range does not have a significant effect on the exciton spectrum.

This paper is organized as follows. In Sec.~\ref{sec:theory} we model the confining potential, introduce the effective mass Hamiltonian and review the relevant physical observables. We describe the confinement scheme setup and discuss the difference in bright and dark excitons. In Sec.~\ref{sec:results} we present the numerical solution of the introduced model Hamiltonian and analyze the oscillator strength of the confined states as a function of the confinement voltage and other physical observables.
In Sec.~\ref{sec:conclusion} we present the main conclusions of our work.

\section{\label{sec:theory}Theory}

\subsection{Effective mass model}

For excitons at the $\mathbf{K^+}/\mathbf{K^-}$ points in TMDs, the effective mass model has proven to be an accurate approach to model the many-body problem\,\cite{Berkelbach2013}.
The effective mass Hamiltonian with an external potential reads
\begin{align}
H = &-\frac{\hbar^2}{2m_\text{e}} \nabla^2_{\mathbf{r}_\text{e}} 
    - \frac{\hbar^2}{2m_\text{h}} \nabla^2_{\mathbf{r}_\text{h}}\\
    &+ V_\text{RK}(|\mathbf{r}_\text{e} - \mathbf{r}_\text{h}|) 
    - V_\text{es}(\mathbf{r}_\text{e}) 
    + V_\text{es}(\mathbf{r}_\text{h}), \nonumber
\end{align}
where $m_\text{e}$ ($m_\text{h}$)~\cite{Thureja2022} is the effective electron (hole) mass and $\mathbf{r}_\text{e}$ ($\mathbf{r}_\text{h}$) describes the position of the electron (hole). The electrostatic in-plane potential is given by $V_\text{es}$. 
The Coulomb interaction in monolayers experiences strong screening effects caused by the surrounding environment of the sample. The Rytova-Keldysh potential~\cite{RK} accurately describes screening for monolayer TMDs\,\cite{Berkelbach2013,Chernikov2014}
\begin{equation}\label{eq:rytova_keldysh}
V_{RK}(r) = -\frac{\pi e^2}{2\epsilon_1 r_0} \left( H_0\left(\frac{r}{r_0}\right) - Y_0\left(\frac{r}{r_0}\right) \right),
\end{equation}
where $H_0$ ($Y_0 $) are the Struve (Bessel) functions of the second kind and  $r_0 = d\epsilon_1/2\epsilon_2$ is the effective screening length,
where $\epsilon_1$ is the dielectric constant of the monolayer, $\epsilon_2$ is the dielectric constant of the material embedding it and $d$ is the thickness of the monolayer.

It is convenient to reformulate the problem in center-of-mass (COM) $\mathbf{R} = (m_\text{e} \mathbf{r}_\text{e}+m_\text{h}\mathbf{r}_\text{h})/M$ and relative coordinates
$\mathbf{r}=\mathbf{r}_\text{e}-\mathbf{r}_\text{h}$,
\begin{equation}
    H = -\frac{\hbar^2}{2M} \nabla^2_{\mathbf{R}} 
    - \frac{\hbar^2}{2\mu} \nabla^2_{\mathbf{r}} 
    + V_\text{RK}(\mathbf{r}) 
    + V_{\text{conf}}(\mathbf{R},\mathbf{r}), 
    \label{eq:COMhamilton}
\end{equation}
with $M=m_\text{e}+m_\text{h}$, $\mu = m_\text{e} m_\text{h}/(m_\text{e}+m_\text{h})$. The confinement of the exciton along the $R_x$ coordinate (see Fig.~\ref{fig:set_up}) arises from the Stark effect. It is therefore convenient to combine the electrostatic potential acting on the electron and hole into the confinement potential $V_{\text{conf}}(\mathbf{R},\mathbf{r}) = -V_\text{es} (\mathbf{R}-\frac{m_e}{M}\mathbf{r})+ V_\text{es} (\mathbf{R}+\frac{m_h}{M}\mathbf{r})$. Due to the translational invariance of $V_\text{conf}$ in the $y$-direction (see Fig.~\ref{fig:set_up}), the exciton wavefunction decouples and can be written as 
\begin{equation} \label{eq:wave_full}
    \Phi_{k_y,n}(\mathbf{R}, \mathbf{r}) = e^{ik_yR_y} \phi_n(R_x, \mathbf{r}),
\end{equation}

\noindent
with energy levels 

\begin{equation}
    E_{k_y,n} = \frac{\hbar^2k_y^2}{2M}+E_n.
\end{equation}

\noindent
Here $E_n$ are the eigenenergies of $\phi_n$. The solution in the $R_y$ direction is a plane wave. The non-trivial part $\phi_n$ and their corresponding energy levels $E_n$ carry the information on the exciton confinement. We solve for these states through an exact diagonalization scheme by mapping Eq.~\eqref{eq:COMhamilton} on a finite sized grid.





The electrostatic potential $V_{\text{es}}$ is obtained through electrostatic simulations of the set up shown in Fig.~\ref{fig:set_up}. A detailed description of the simulations can be found in the Appendix section~\ref{sm:COMSOL}.
The TMD monolayer is sandwiched between two layers of h-BN with electrical gates of different lengths on the top and the bottom.
Applying opposite bias ($V_\text{BG}>0, V_\text{TG}<0$) creates
a potential difference between the regions covered by both gates and the region only covered by the bottom gate. 

\subsection{Physical observables}\label{observ}
The exciton's oscillator strength characterizes how well it couples to light and it is governed by the transition rate $\Gamma$ of a photon exciting an electron from the valence into the conduction band. Using Fermi's golden rule,
\begin{equation}
    \Gamma =\left( \frac{2\pi}{\hbar} \right) 
\sum_n \left| \bra{f_n} H_\gamma \ket{0} \right|^2 
\delta\left(E_n(K_x) - E_0 - \hbar \omega\right).
\end{equation}
Here $\ket{f_n}$ are the exciton states and $E_n(K_x)$ are the corresponding energies of the exciton with wavevector \textbf{K}, $H_\gamma$ is the exciton-photon interaction and 
$\ket{0}$ is the ground state with no excitons present. 
$f_n$  can be written as
\begin{equation}\label{eq:basis_decomposition}
    f_n = \sum_{R_x}\sum_{\mathbf{r},\,\mathbf{k}} \frac{1}{\sqrt{N}} \phi_n(R_x,\mathbf{r})  u_{\mathbf{k}}^\text{c}(\mathbf{r}_\text{e})  u_{\mathbf{-k}}^\text{v}(\mathbf{r}_\text{h}),
\end{equation}
where $N$ is the number of unit cells the exciton extends over. 
$u_\mathbf{k}(\mathbf{r}_\text{e})$ ($u_{\mathbf{-k}}(\mathbf{r}_\text{h})$) are the periodic parts of the Bloch functions for electrons (holes) with $\text{c}$ $(\text{v})$ indicating the electron (hole) being in the conduction (valence) band. $\phi_n(\,\mathbf{r},R_x)$ is the envelope wavefunction given by Eq.~\eqref{eq:wave_full}. We assume that $H_\gamma$ only acts on the electronic states $u_\mathbf{\pm k}(\mathbf{r}_\text{e/h})$.
Using Eq.~\eqref{eq:basis_decomposition} the matrix element $\braket{f_n | H_\gamma | 0} $ is written as
\begin{align}\label{equation:matrix}
\braket{f_n | H_\gamma | 0} 
=&\sum_{R_x}\sum_{\mathbf{r},\mathbf{k}}    \frac{1}{\sqrt{N}} e^{i \mathbf{k} \cdot    \mathbf{r}} \phi_{n}(R_x,\mathbf{r})  \\ \nonumber
&\braket{u_\mathbf{k}^\text{c}(\mathbf{r}_\text{e})u_{\mathbf{-k}}^\text{v} (\mathbf{r}_\text{h})    | H_\gamma|     0} \nonumber \\ \nonumber
=& \sum_{R_x}\sum_{\mathbf{r},\mathbf{k}} \frac{1}{\sqrt{N}} e^{i \mathbf{k} \cdot \mathbf{r}} \phi_{n}(R_x,\mathbf{r}) \\ \nonumber
&\braket{u_\mathbf{k}^\text{c}(\mathbf{r}_\text{e})| H_\gamma | u_\mathbf{k}^\text{v}(\mathbf{r}_\text{h})},
\end{align}
Due to the negligible photon momentum we can approximate $\braket{u_{\mathbf{k}}^\text{c} | H_\gamma | u_{\mathbf{k}}^\text{v}}$ to be independent of $\mathbf{k}$, therefore the sum over $\mathbf{k}$ simplifies to a delta function $\delta (\mathbf{r})$, which makes the summation over $\mathbf{r}$ trivial. In the continuum limit the summation over $R_x$ is replaced by an integral
\begin{align}\label{eq:osc_strength}
    |\braket{f_n | H_\gamma |0}|^2 = &\Big| \int dR_x\phi_{n}(R_x, \mathbf{r}=0)\Big|^2 \\ \nonumber
    &\Big |\braket{u_\mathbf{k}^\text{c}(\mathbf{r}_\text{e})| H_\gamma | u_\mathbf{k}^\text{v}(\mathbf{r}_\text{h})}\Big |^2.
\end{align}
Here, $\phi_{n}(R_x, \mathbf{r}=0)$ quantifies the spatial overlap of the electron and hole envelope function at $R_x$, which is related to 
the probability of finding the electron and hole on the same lattice site. In experiments~\cite{Thureja2022,thureja2024,Heithoff} the optical activity of the confined  states is measured by the reflectance of incident light relative to the the reflective properties of the unconfined exciton. Therefore we can also define the relative oscillator strength as

\begin{equation}\label{relative}
    F_n = \frac{\braket{f_n | H_\gamma | 0} }{\braket{f_\text{2D} | H_\gamma | 0} } = \frac{|\int dR_x\phi_{n}(R_x, \mathbf{r}=0)|^2}{|\int dR_x \phi_{\text{2D}}(R_x, \mathbf{r}=0)|^2},
\end{equation}
where $\Phi_\text{2D}$ is the unconfined 2D exciton state. 

In what follows we use the notation $\braket{\hat{A}}_n$ with
\begin{equation}
\braket{\hat{A}}_n = \int\int d\mathbf{r} dR_x \, \phi_n(R_x, \mathbf{r})^\dagger \hat{A}(R_x, \mathbf{r}) \phi_n(R_x, \mathbf{r})
\end{equation}
to represent the expectation value of a given observable $\hat{A}$ for the $n$-th eigenstate.

\subsection{Dark States}\label{sub:dark}

Exciton states with COM-momentum outside of the light cone are not accessible by photon excitation. Additionally, there are spin-forbidden states where the excitation of the electron into the conduction band requires a spin-flip. These states are commonly refered to as dark states~\cite{Selig2016,Poem2010,Tseng2015,Roszak2007,Ye2014}. Here, we introduce a new variant of the dark state, which arises as a result of symmetry-forbidden states. Considering the integrand in Eq.~\eqref{eq:osc_strength}, we find that the oscillator strength is 0 when the envelope function $\phi_n(R_x,\mathbf{r}=0)$ has odd parity. Hence, these states are not accessible by photon excitation and will be referred to as dark states in the following.

\section{Results}\label{sec:results}

\begin{figure}[htpb]
    \centering
    \includegraphics[width = \linewidth]{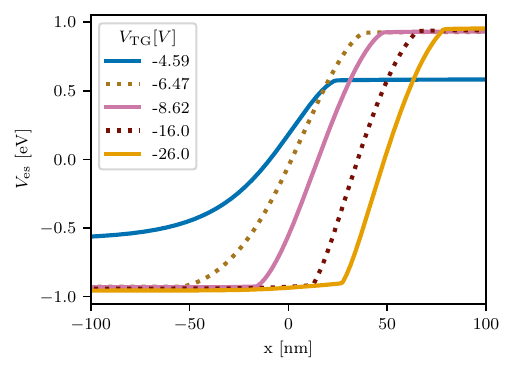} 
    \caption{ Subset of in-plane electrostatic potentials $V_{\text{es}}$ obtained through COMSOL simulations of the setup shown in Fig.~\ref{fig:set_up}, with an hBN thickness of 30nm per slab. For convenience the potential is set to $V_\text{es}(x_0)=0~\si{\eV}$, where $x_0$ is the midpoint of the step. }
    \label{fig:potential}
\end{figure}

For our simulations we consider the set up shown in Fig.~\ref{fig:set_up} with $\text{MoSe}_2$ as the TMD monolayer, which is identical to the set up used in Ref.~\cite{Thureja2022}. For additional information we also refer to Appendix~\ref{sm:COMSOL}. The bottom gate is kept at a constant $V_\text{BG}=4$V, while the top gate voltage $V_\text{TG}$ is treated as an adjustable parameter to tune the height of the potential step. In our simulations we probe the $V_\text{TG}$ range $[-32\si{\V},0\si{\V}]$, which exceeds the experimental range of $[-10\si{\V},0\si{\V}]$ presented in Ref.~\cite{Thureja2022}. This serves to probe the convergence behavior of the confined exciton spectrum at large voltage biases. 
A selection of simulated in-plane potentials is shown in Fig.~\ref{fig:potential}. Increasing the bias-voltages at the electrical gates leads to a build up of doping charge in the areas covered by the respective gates. This build up of charge carriers screens the electric potential created by the gates. 
Therefore we observe for large $|V_\text{TG}|$ that the shape of the potential step saturates. Further increases in voltage only shifts the center of the potential step in the positive x-direction. 

\begin{figure}[htbp]
    \centering

    \includegraphics[width=\linewidth]{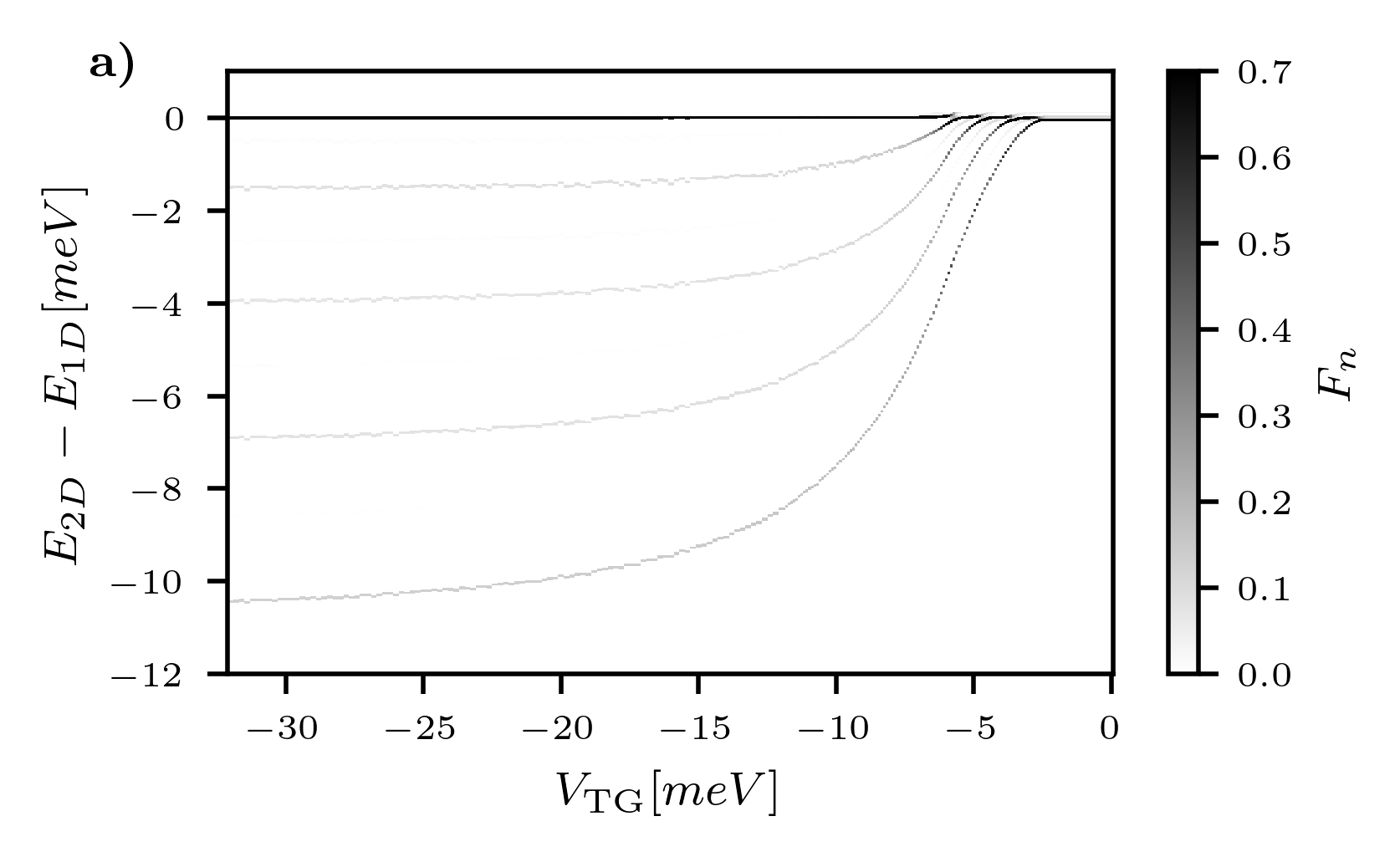}
    \includegraphics[width=\linewidth]{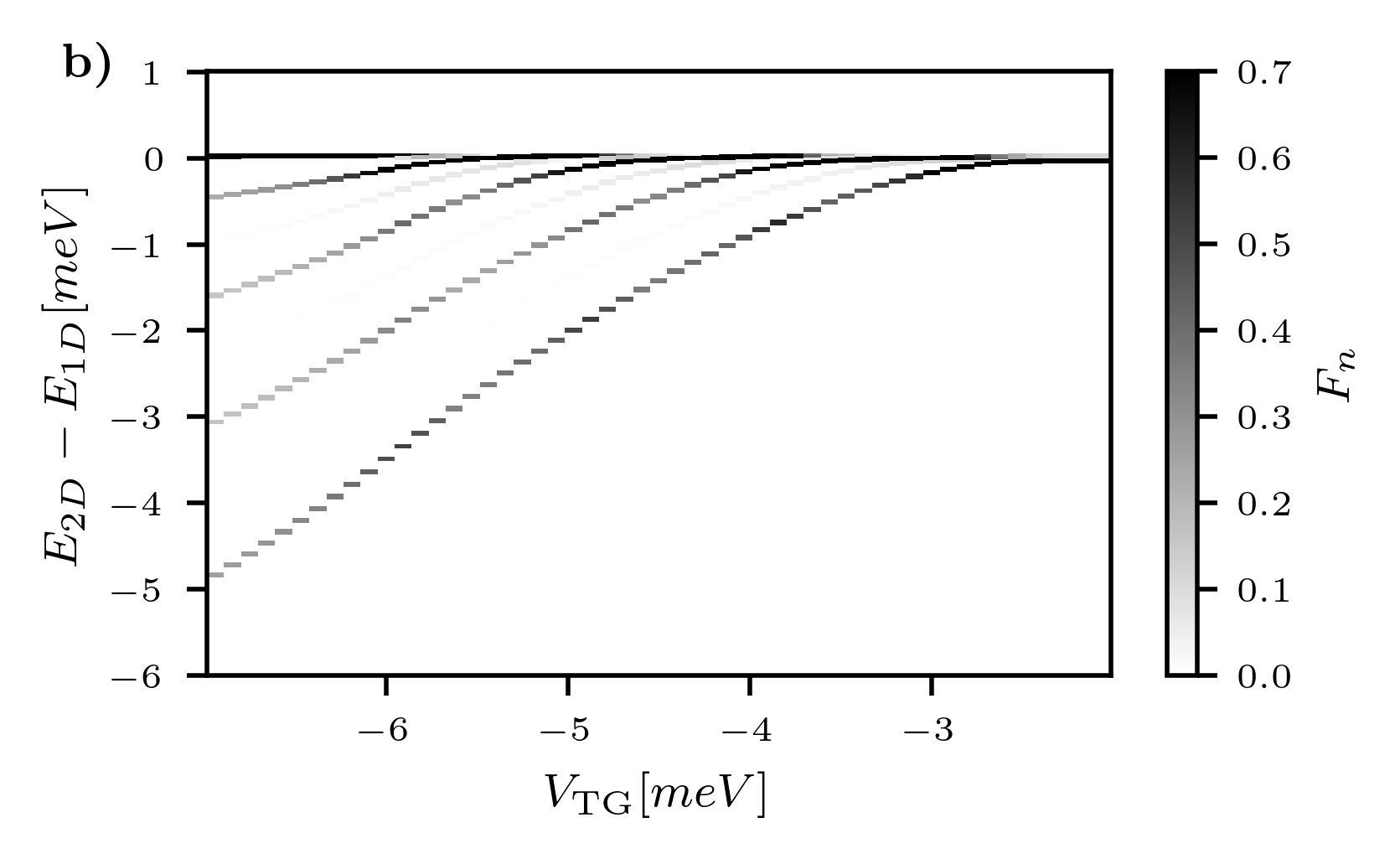}

        \caption{\textbf{a)}  Oscillator strength and corresponding energy of the confined states relative to the unconfined 2D state for a $V_\text{TG}$ range of $[-32\si{\V},0\si{\V}]$. The four emerging discrete lines are the bright states with even $n$. The energy levels of the states converge to constant values for large $V_\text{TG}$, due to the minimal changes in the confining potential in the large voltage regime, as seen in Fig.~\ref{fig:potential}. \textbf{b)} Zoom into the region of $V_\text{TG} \in [–2\si{\V},-7\si{\V}]$. We observe the emergence of the odd states with faint oscillator strength. The linewidths in these plots have no physical meaning and are purely for the purpose of visibility, while each horizontal line represents a single data point in our parameter space their length is also arbitrarily chosen for visibility.}
    \label{fig:osc_strength}
\end{figure}

Fig.~\ref{fig:osc_strength} \textbf{a)} shows the relative oscillator strength $F_n$ of the confined exciton states and the energy of the unconfined exciton in the 2D limit with their respective eigenenergies as a function of $V_\text{TG}$. As a result of the potential saturating for large $|V_\text{TG}|$, we find that the exciton spectrum also saturates, with a progressively declining redshift of the confined exciton states.
For the unconfined 2D exciton, $R_x$ and $\mathbf{r}$ decouple and we can obtain the binding energy from the relative problem. We find a binding energy of $E_\text{b,2D} = \SI{221.2}{\milli\electronvolt}$, which is in good agreement with the binding energies found in Ref.~\cite{Goryca2019}. The unconfined state is shown in Fig.~\ref{fig:osc_strength} as the horizontal black line at 0. 
In Fig.~\ref{fig:osc_strength} \textbf{b)} we plot a zoom-in for the region of $V_\text{TG} \in [-2\si{\V},-7\si{\V}]$. Here we observe that there are states with significantly smaller oscillator strength in between the four aforementioned states. These are the dark states as we described in Sec.~\ref{sub:dark}. The small but finite oscillator strength is a result of the asymmetrical confining potential leading to envelope functions $\phi_n(R_x,\mathbf{r}=0)$ that do not have perfect odd parity. We note that the states with even (odd) parity also have even (odd) quantum number $n$. In the following these states will be referred to as bright (dark) states.

\begin{figure}[htb]
    \centering
    \includegraphics[width=\linewidth]{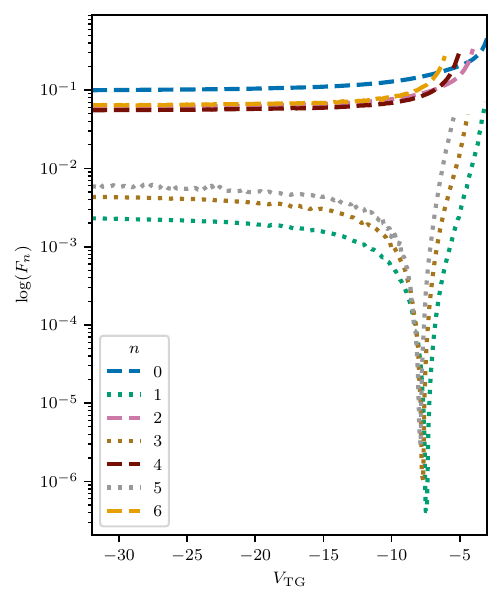}
    \caption{  Relative oscillator strength (logarithmic scale) for the bright (dashed) and dark states (dotted). The bright states have oscillator strengths that are an order of magnitude or more stronger than the dark states. The significant difference is explained by the parity of the corresponding wavefunctions. The dip at $V_\text{TG}=8.62$V is explained by the symmetry of the confining potential around the mid-point of the step, which corresponds to $F_n=0$ for odd n. The symmetric confining potential (shown in Fig.~\ref{fig:potential}) is caused by equivalent doping densities in the p-doped and n-doped region.}
    \label{fig:KN}
\end{figure}

A comparison of the oscillator strength between bright and dark states as a function of $V_\text{TG}$ is shown in Fig.~\ref{fig:KN}. The oscillator strength of the bright states is generally an order of magnitude or more larger than their dark counterparts. Furthermore, we observe that the oscillator strength for the dark states drops to 0 at $V_\text{TG} = -8.62\si{\volt}$. This is a result of the electrostatic potential approaching a perfectly symmetric form around its own midpoint at that top-gate voltage, which therefore results in an envelope function $\phi_n(R_x,\mathbf{r}=0)$ with exact odd parity. 

\begin{figure}
    \centering
    \includegraphics{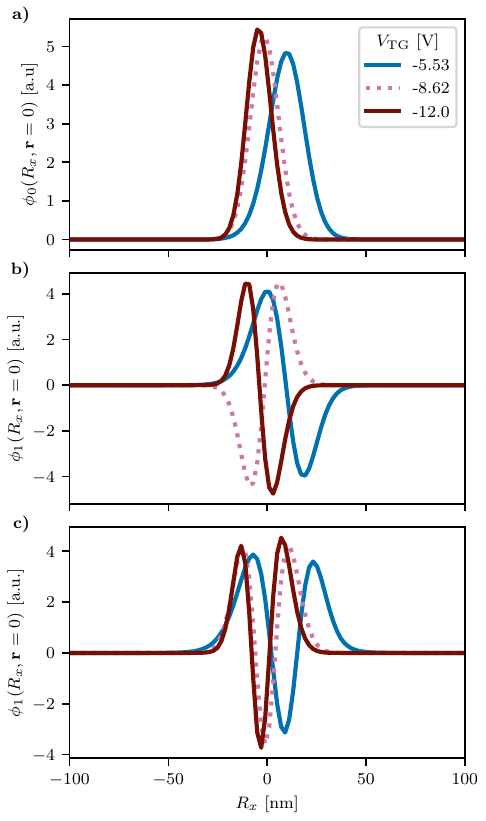}
    \caption{\textbf{a)}-\textbf{c)}: The wavefunctions $\phi_n(R_x, \mathbf{r}=0)$ for selected confining potentials. These functions also act as the integrand in Eq.~\eqref{relative}. Parts with opposite sign in the wavefunction reduce the oscillator strength. In the case of odd $n$ this leads to a significant reduction in oscillator strength leading to the dark states. For $V_\text{TG} = -8.62\si{\volt}$, $\phi_n(R_x, \mathbf{r}=0)$ possesses perfect odd parity and leads to the dip in oscillator strength shown in Fig~\ref{fig:KN}. }
    \label{fig:integrand}
\end{figure}

While the parity determines the categorization into bright and dark states, we find that the number of nodes in $\phi_n(R_x,\mathbf{r}=0)$ is also a main contributing factor for the oscillator strength of the bright states at large $|V_\text{TG}|$. The envelope function for the bright state with $n=0$ and $n=2$ for selected $V_\text{TG}$ is shown in Fig.~\ref{fig:integrand} \textbf{a)} and \textbf{c)} respectively. Considering the integration in Eq.~\eqref{relative}, we find that parts of the envelope function with opposite sign will counteract in the integration and therefore lead to a smaller oscillator strength. Hence, the ground-state with no nodes has the largest oscillator strength at large $|V_\text{TG}|$. Additionally, the quantum number $n$ is also the number of nodes in $\phi_n(R_x,\mathbf{r}=0)$. This means that the number of optically accessible states is limited for any arbitrary potential setup, due to the increasing number of nodes leading to decreasing oscillator strength in the bright states. 
The second  contributing factor is the confinement length. With tighter confinement of the exciton wavefunction, the envelope functions reduce in width and therefore the integrand in Eq.~\eqref{relative} is also compressed. Practically, this reduces the area under the curve and leads to a smaller oscillator strength for tighter confinement. This is the dominating effect for the confined states as they emerge out of the 2D continuum in Fig.~\ref{fig:KN}. In the following we are going to focus our analysis on the bright states. 

\begin{figure}
    \centering
    \includegraphics{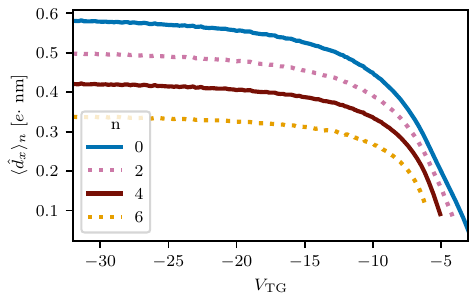}
    \caption{Electrical dipole moment $\braket{\hat{d}_x}_n$ of  the bright states. The dipole in the $r_y$-direction is 0, due to the translational invariance of the confining potential in $r_y$-direction. The exciton in the 2D limit has no electric dipole.}
    \label{fig:dipole}
\end{figure}

The electric field induces an electric dipole moment $q\braket{\hat{\vec{r}}}_n$ in the exciton.
The $r_x$-component of the dipole moments $q\braket{\hat{r}_x}_n$ for the bright states is shown in Fig.~\ref{fig:dipole}. Since the confining potential is invariant under translation in the $r_y$-direction, $q\braket{\hat{r}_y} = 0$.  We observe that larger $|V_\text{TG}|$ are associated with larger $\vec{d}$, as the steeper confinement potential induces larger electrical dipole moments. With increasing $|V_\text{TG}|$ it becomes energetically more favorable for the electron (hole) to be closer to the n-doped (p-doped) region. Therefore the dipole moment increases for increasing magnitude in $V_\text{TG}$.

\begin{figure}[htbp]
    \centering
    \includegraphics{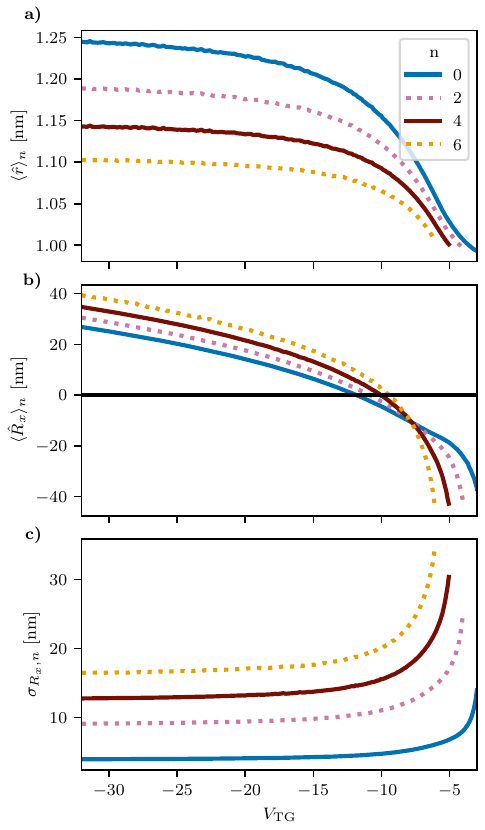}
    \caption{\textbf{a)} Exciton radius as a function of $V_\text{TG}$. The radius increases from 0.96nm for the unconfined 2D limit up to 1.24nm (ground state) in the limit of large voltages. Larger radii are energetically more favorable, due to the confining potential pulling the exciton apart.  \textbf{b)} $\braket{\hat{R}_x}$ as a function of $V_\text{TG}$. \textbf{c)} Confinement length given by the standard deviation $\sigma_{R_x}$. The lower lying states are subject to a tighter confinement. }
    \label{fig:position}
\end{figure}

The same effect is seen in Fig.~\ref{fig:position} \textbf{a)} where we show that the exciton radius also increases with the magnitude of $V_\text{TG}$, with the radius of the ground state increasing from 0.96nm at $V_\text{TG}=0$V to 1.24nm at $V_\text{TG}=-32$V. This confirms that the spatial extent of the exciton increases for increasing $|V_\text{TG}|$. 
However, this does not agree with the measurements of energy splits under magnetic fields from Ref.~\cite{Thureja2022}, where they estimate a root mean square of $\sqrt{\braket{r^2}}\approx6\si{\nm}$ under the assumption of a simple hydrogenic picture for the exciton states. This discrepancy further underlines the assumption of Thureja et al. that a simple hydrogenic approach to the wavefunction of the confined exciton is insufficient. 

We now turn to the behavior of the COM coordinate as a function of confining potential. The relation between $\braket{\hat{R}_x}$ and the top gate voltage is shown in Fig.~\ref{fig:position} \textbf{b)}. We observe that the bright states are centered around different positions in space. 
Similar to the symmetry discussion on the oscillator strength  $F_n$ for the dark states, we observe a crossover point at $V_\text{TG}= -8.62$V, where $\braket{\hat{R}_x}$ is the same for all bright states due to the aforementioned symmetry of the potential. Additionally, $\braket{\hat{R}_x}$ increases for large $|V_\text{TG}|$ as a result of the $V_\text{es}$ midpoint as shown in Fig.~\ref{fig:potential}.
In Fig.~\ref{fig:position} \textbf{b)} the standard deviation of $\braket{\hat{R}_x}$ is shown as $\sigma_{R_x,n}$ for the $n$-th state, which we interpret as the confinement length. We observe that the energetically more favorable states are subject to smaller confinement lengths. Furthermore, the confined states exhibit the same convergence behavior for their confinement length as the oscillator strength $F_n$, due to the previously discussed convergence in the potential. 

\begin{figure*}[htbp]
  \centering
      \includegraphics{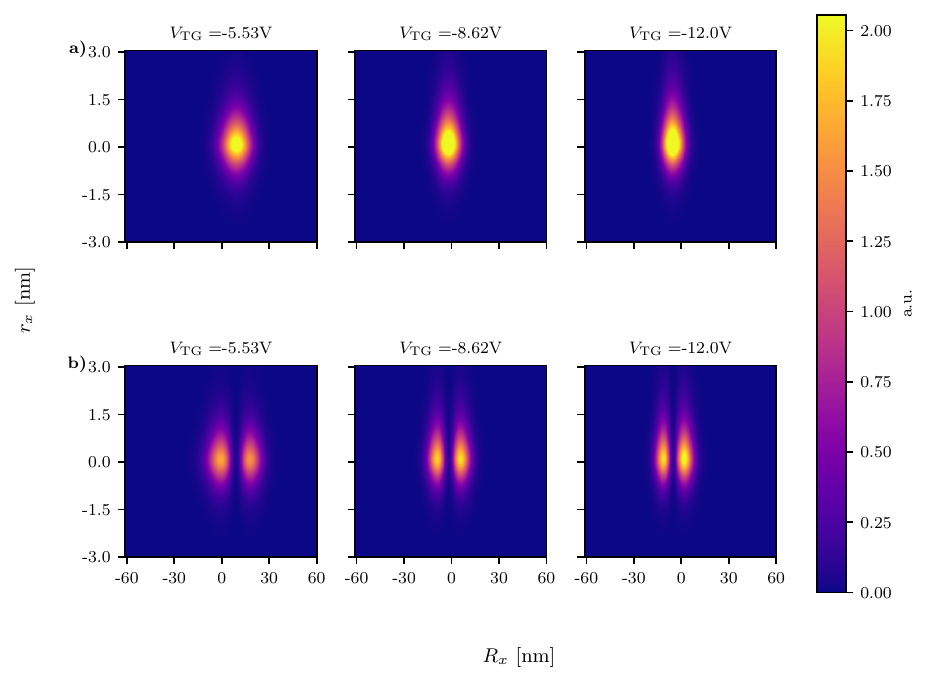}
  \caption{The probability density $|\phi _n(R_x , r_x, r_y = 0)|^2$ for selected $V_\text{TG}$ for $n=0$ in \textbf{a)} and $n=1$ in \textbf{b)}. For increasing $|V_\text{TG}|$ we observe increasing spatial extent in the $r_x$ coordinate and decreasing spatial extent in the $R_x$ coordinate. The probability densities show that the description of the confined exciton states based on a simple hydrogenic model is insufficient.}
  \label{fig:wavefunction}
\end{figure*}

Complementary, we show the probability density $|\phi(R_x,r_x,r_y=0)|^2$ for selected confinement voltages in Fig.~\ref{fig:wavefunction}. The confinement of the exciton is visible as well as the increased separation of the electron and hole due to the prominent tail in the positive $r_x$ direction. In combination with the non trivial shape of $\phi_n(R_x,\mathbf{r})$ as a function of $R_x$, this provides further proof that the hydrogenic picture is insufficient to capture the physics in confined exciton states.  

\section{Summary and conclusion}\label{sec:conclusion}
We have studied  excitons confined to 1D that arise in monolayer TMDs when an in-plane electric field is applied through electrical gates. We employed an exact diagonalization method by mapping the effective mass Hamiltonian to a finite sized grid. This includes a calculation of the entire confined exciton spectrum over a voltage regime $V_\text{TG}\in [-32\si{\V},0\si{\V}]$ that exceeds the range found in experimental setups. 
The numerical study shows the emergence of up to seven confined exciton states, where the number of confined states is determined by the steepness of the potential step. The confined states $\phi_n(R_x,\mathbf{r})$ can be classified into states with even and odd parity in $\phi_n(R_x,\mathbf{r})$ which we call bright and dark respectively, due to their difference in oscillator strength. We find that the dark states carry oscillator strengths that are at least an order of magnitude lower than their bright counterparts, and can be completely dark for highly symmetric confining potentials. Due to this difference in oscillator strength only the bright states are detected in Ref.~\cite{Thureja2022}. 
The bright states found in our numerical study are in agreement with the experimentally measured spectrum. Our calculations do not incorporate the many-body interactions of the exciton with the present charge doping, therefore a comparison with Ref.~\cite{Thureja2022} is only possible in a $V_\text{TG}$ range limited by the repulsive polaron branches. For large $|V_\text{TG}|$ the steepness of the potential step converges due to the buildup of excess charge carriers in the areas covered by the electrical gate leading to screening. As a consequence, going beyond the experimentally available parameter range of $[-10\si{\V},0\si{\V}]$~\cite{Thureja2022} does not result in significant changes in the confined exciton spectrum. 
Additionally, we find that the applied electric potential induces an electrical dipole in the confined exciton which also results in lower oscillator strength, due to the reduced electron-hole overlap. 
Our results provide an insightful theoretical background to the experimental results presented in Ref.~\cite{Thureja2022}. The discussed dark states with small but finite oscillator strength may facilitate the research of further confinement schemes with the ability to access the dark states optically. Furthermore, the presented insights on the induced electric dipoles paves the way for future electric manipulation schemes for the confined excitons.

\begin{acknowledgments}
    We acknowledge funding from the Deutsche Forschungsgesellschaft (DFG) through the Matter and Light for Quantum Computing (ML4Q) Excellence Cluster EXC 2004/1-390534769. The numerical calculations were perfomed on computational ressources provided by the RWTH Aachen University (project ID: RWTH1610).
\end{acknowledgments}


\clearpage
\bibliographystyle{apsrev4-2}
\bibliography{output}

\clearpage
\section*{Appendix} \label{sec:supplementary}

\section{Electrostatic simulation}\label{sm:COMSOL}

We simulate the electrostatic potential $V_{\text{es}}(x, z)$ in the heterostructure with a finite-element method COMSOL, following the simulations presented in the Appendix of Ref.~\cite{Thureja2022}. $V_{\text{es}}(x, z=0)$ denotes the potential at the position of the semiconductor. We define the sample geometry as seen in Fig.~\ref{fig:set_up}. The monolayer is encapsulated by $\SI{30}{\nano\meter}$ of h-BN from both sides. In the simulation it extends over $\SI{600}{\nano\meter}$ in lateral direction to suppress edge effects while not being computationally expensive. We assign the dielectric constants of h-BN to the in-plane $\epsilon_\parallel $ and out-of-plane $\epsilon_\perp$ axes of the geometry, see Tab.~\ref{tab:comsol_parameters}. We assume metallic gates that are homogeneously biased. Hence, we define the bottom and top gates by setting electrostatic boundary conditions, with $V_{\text{BG}}$ and $V_{\text{TG}}$, respectively. For all remaining outer edges of the structure we use the boundary condition $\vec{n} \cdot \vec{D} = 0$, with $\vec{n}$ being the normal vector and $\vec{D}$ the electric flux density such that no charge is present. 

\begin{table}[b]
\caption{
Parameters used for the simulation. $m_0 $ is defined as the mass of the electron. }
\label{tab:comsol_parameters}
\begin{ruledtabular}
\begin{tabular}{lcc}
Parameter & Value & Cite\\
\hline
$E_\text{F} - E_\text{VB}$  &$\SI{0.99}{\electronvolt}$&~\cite{valence_edge} \\
$E_\text{G}$   & $\SI{1.85}{\electronvolt}$& \cite{Thureja2022}\\
$m_\text{e}$ &  $ 0.7\,m_0$ &\cite{Larentis} \&~\cite{Goryca2019}\\
$m_\text{h}$ & $ 0.6\,m_0$&~\cite{Goryca2019}\\
$\epsilon_\perp$ & $3.74$ &\cite{Laturia2018}\\
$\epsilon_\parallel$ & $6.93$ &\cite{Laturia2018}\\
\end{tabular}
\end{ruledtabular}
\end{table}

We allow for a surface charge density at the semiconductor plane by using the built-in function that ensures a self consistent calculation of the electrostatic potential including charges on the semiconductor. The resulting charge carrier distribution $\sigma(x)$ is altered by the applied voltages $V_{\text{BG}}$ and $V_{\text{TG}}$ and is given by the density of electrons $\sigma_\text{n}(x)$ and holes $\sigma_\text{p}(x)$ 
\begin{equation*}
 \sigma(x) = \sigma_\text{p}(x) + \sigma_\text{n}(x).
\end{equation*}
We use the Thomas-Fermi approximation to describe the individual densities as
\begin{align}
\sigma_\text{n}(x) &= - e \int_{E_\text{CB}(x)}^{E_{\text{F}}} \mathcal{D}_\text{n}(E) dE, \quad E_\text{F}>E_\text{CB}(x)\\
&= - e\, \mathcal{D}_\text{n}(E)(E_\text{F}-E_\text{CB}(x))
\end{align}
and
\begin{align}
 \sigma_\text{p}(x) &= e \int_{E_{\text{F}}}^{E_\text{VB}(x)} \mathcal{D}_\text{p}(E) dE, \quad E_\text{F}<E_\text{VB}(x)\\
 &= e \,\mathcal{D}_\text{p}(E)(E_\text{VB}(x)-E_\text{F}).
\end{align}
Here $\mathcal{D}_{i}(E)$ is the 2D density of states for electrons and holes in the TMD and given by
\begin{equation}
    \mathcal{D}_i(E) = \frac{g_s g_v m_i}{2\pi \hbar^2},
\end{equation}
where $g_\text{s}= 1$ is the spin degeneracy, $g_\text{v}=2$ is the valley degeneracy, and $m_i$ is the effective mass of the electron or hole. $E_\text{CB}(x)$ and $E_\text{VB}(x)$ denote the conduction and valence band energies and $E_{\text{F}}$ is the Fermi energy which is determined by the contact (see Tab.~\ref{tab:comsol_parameters} for used values). Applying voltages to the gates shifts the energetic position of the band edges depending on the local electrostatic potential as
\begin{equation}
     E_{j} (x) = E_{j} - V_{\text{es}}(x,z=0), \quad j = \text{CB}, \text{VB}.
\end{equation}

\begin{figure}[htpb]
    \centering
    \includegraphics[width=\linewidth]{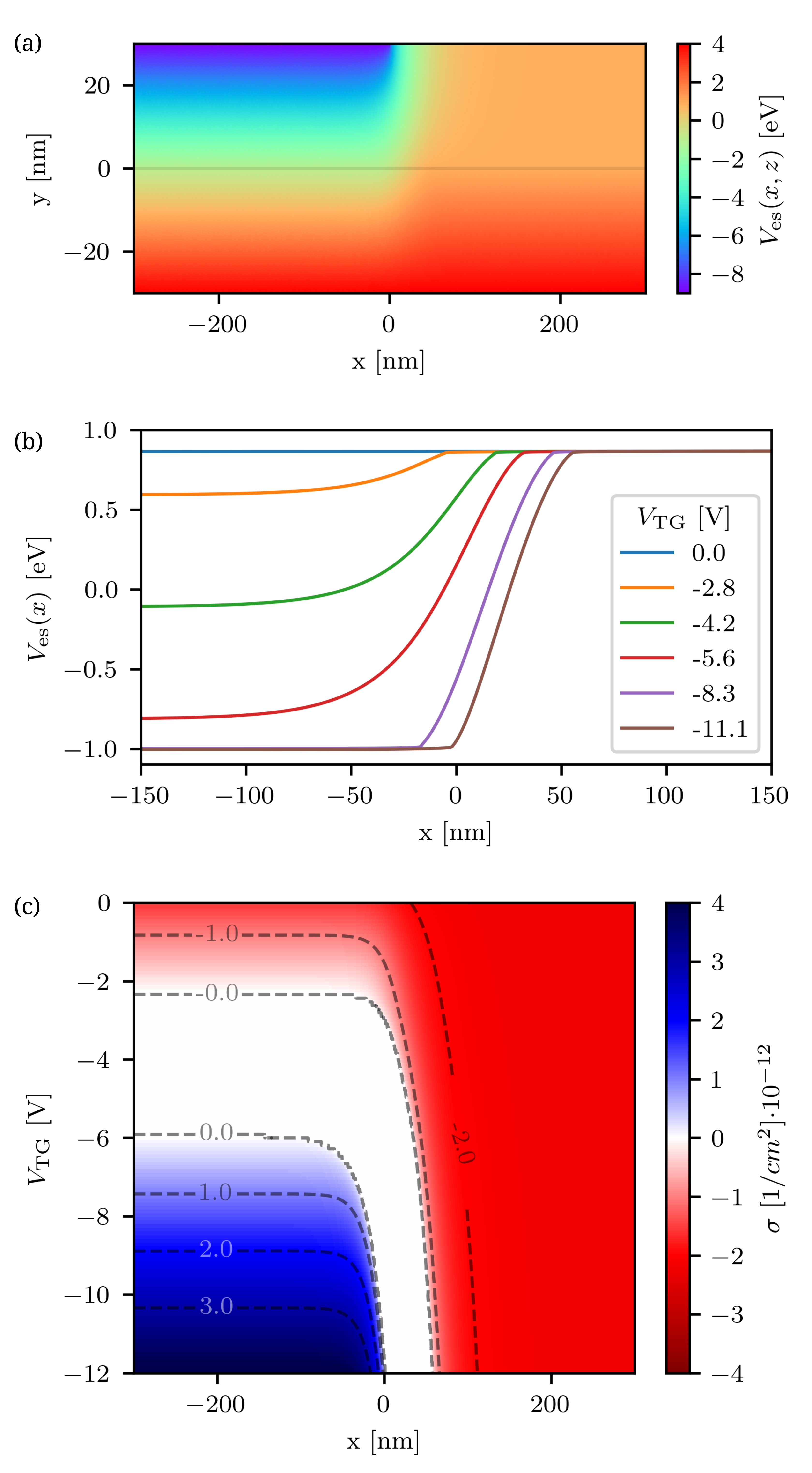}

    \caption{Electrostatic potential based on simulation of the sample for $V_\text{BG}=\SI{4}{\volt}$. a) $V_\text{es}(x,z)$ shown for the sample cross-section with the top gate at $x\leq0$ and the gate voltages $V_\text{TG}=\SI{-9}{\volt}$.   
    The TMD is positioned at $z = \SI{0}{\nano\meter}$
    (grey line). b) Electrostatic potential in the plane of the TMD ($z=0$) for different top gate voltages $V_\text{TG}$. c) Charge carrier density $\sigma(x)$ at the TMD for tuned top gate voltage. Red (blue) indicates electron (hole) doping.
    }
    \label{fig:electrostatics}
\end{figure}

Figure~\ref{fig:electrostatics} a) shows the electrostatic potential $V_{\text{es}}(x, z)$ in the crosssection of the sample calculated for $V_\text{BG}=\SI{4}{\volt}$ and $ V_\text{TG}=\SI{-9}{\volt}$. Tracking the potential along the plane of the TMD ($z=0$) reveals a step in $V_\text{es}(x,z=0)$, shown in Fig.~\ref{fig:electrostatics} b), that becomes sharper for increasing $V_\text{TG}$. For $E_F > E_\text{CB}(x)$ ($E_F < E_\text{VB}(x)$), the semiconductor shows electron (hole) doping. This results in the saturation of $V_{\text{es}}(x,z=0)$ at values corresponding to the valence and conduction band energies as the charges compensate the local in-plane electric field $F_x (x)$. The charge carrier density is shown in Fig.~\ref{fig:electrostatics} c) for tuned top gate voltage, illustrating the transition from an electron-doped sample to a p-i-n junction with a narrow intrinsic region.

\end{document}